\documentstyle[12pt]{article}
\newcommand{\Ref}[1]{(\ref{#1})}
\newcommand{\C}{{\bf C}}
\newcommand{\s}[3]{\sigma_{#1}({#2},{#3})}
\newcommand{\R}[2]{R_{#1#2}(\xi_{#1#2})}
\newcommand{\Z}{{\bf Z}}
\newcommand{\ben}{\begin{equation}}
\newcommand{\een}{\end{equation}}
\newcommand{\bean}{\begin{eqnarray}}
\newcommand{\eean}{\end{eqnarray}}
\newcommand{\be}{\begin{displaymath}}
\newcommand{\ee}{\end{displaymath}}
\newcommand{\bea}{\begin{eqnarray*}}
\newcommand{\eea}{\end{eqnarray*}}
\newcommand{\half}{{1\over 2}}
\newenvironment{proof}{\noindent{\em Proof\/}:}{$\;\Box$}

\newtheorem%
{thm}{Theorem}
\newtheorem%
{proposition}[thm]{Proposition}
\newtheorem%
{lemma}[thm]{Lemma}
\newtheorem%
{corollary}[thm]{Corollary}
\title{A simple construction of elliptic $R$-matrices}
\author{Giovanni Felder${}^*$ and Vincent Pasquier${}^{**}$}
\date{2 February 1994}
\begin{document}
\maketitle
\centerline{${}^*$Department of Mathematics, University of North Carolina
at Chapel Hill,}
\centerline{Chapel Hill, NC 27514, USA}
\smallskip
\centerline{${}^{**}$Service de Physique Th\'eorique, CEN-Saclay,}
\centerline{91191 Gif-sur-Yvette, France}
\begin{abstract}
We show that Belavin's solutions of the quantum Yang--Baxter equation
can be obtained by restricting an infinite $R$-matrix to suitable
finite dimensional subspaces. This infinite $R$-matrix is a modified
version of the Shibukawa--Ueno $R$-matrix acting on functions of two
variables. (hep-th/9402011)
\end{abstract}

\noindent Shibukawa and Ueno \cite{SU}
have defined an elliptic
$R$-operator acting on the space of
functions of two variables on the circle, and obeying the
quantum Yang--Baxter equation, extending the work of
Gaudin \cite{G}. It has a very simple
form. To describe it let us introduce the basic function $\s wz\tau$
uniquely characterized by having the following behaviour as a function
of $z$ for fixed (generic) $w\in\C$ and $\tau\in\C$, Im$(\tau)>0$:
\bean\label{sig}
\s w{z+1}\tau&=&\s wz\tau,\nonumber\\
\s w{z+\tau}\tau&=&e^{2\pi iw}\s wz\tau,
\eean
and being meromorphic with only simple poles on the lattice
$\Z+\tau\Z$, and unit residue at the origin.

In terms of Jacobi's theta function
\be
\vartheta_1(z,\tau)
=-\sum_{n\in\Z+\half}e^{i\pi n^2\tau+2\pi in(z+\half)},
\ee
we can express $\s wz\tau$ as
\ben\label{sigthe}
\s wz\tau={\vartheta_1(z-w,\tau)\vartheta_1'(0,\tau)\over
           \vartheta_1(z,\tau)\vartheta_1(-w,\tau)},
\een
where the prime means derivative with respect to the
first argument.

The Shibukawa--Ueno $R$-operator is then
\be
R(\xi)f(z_1,z_2)=\s\mu{z_{12}}\tau
f(z_1,z_2)-\s\xi{z_{12}}\tau f(z_2,z_1)
\ee
Here, and below, we make use of the abbreviation
$z_{12}=z_1-z_2$.
The operator $R(\xi)$ maps the space of, say, continuous
1-periodic functions
of $z_1$, $z_2$ to itself for each (generic) value of the
spectral parameter $\xi\in\C$ and ``anisotropy'' parameter $\mu\in\C$.

The main property of $R$, proved in \cite{SU},
is that it obeys the quantum Yang--Baxter
equation (QYBE). If we define $R_{ij}$, $i\neq j$,
 to be $R$ acting on a function
of $n$ variables by viewing it as a function of the  $i$th
and $j$th variable,
then the QYBE is the relation on the space of functions of
three variables
\be
\R 12\R 13\R 23=\R 23\R 13\R 12,
\ee
where, again, $\xi_{ij}=\xi_i-\xi_j$.

Let us now introduce the space of shifted theta functions
of degree $k=1$, 2,\dots : define $V_k(\xi)$ as the
space of entire functions $f$ of one complex variable
such that
\bean\label{theta}
f(z+1)&=&f(z)\nonumber\\
f(z+\tau)&=&\alpha_k(z,\xi)f(z)\\
\alpha_k(z,\xi)&=&e^{-2\pi ikz-\pi ik\tau+2\pi i\xi}\nonumber
\eean
It is well-known  that
$V_k(\xi)$ has dimension $k$. A basis
will be given explicitly below. For $\xi_1$, $\xi_2\in\C$ we identify
$V_k(\xi_1)\otimes V_k(\xi_2)$ with the space of entire functions of
variables $z_1$, $z_2$ belonging to $V_k(\xi_i)$ as functions
of $z_i$, for any fixed value of the other argument.

\begin{proposition}
$R(\xi_{12})$ maps
$V_k(\xi_1)\otimes V_k(\xi_2+\mu)$ to
$V_k(\xi_1+\mu)\otimes V_k(\xi_2)$.
\end{proposition}

\begin{proof}
It is sufficient to show that if $f\in
V_k(\xi_1)\otimes V_k(\xi_2+\mu)$, then $R(\xi_{12})f$
is entire in both variables and has the required properties
under lattice translations.

It is first easy to see, using the behaviour of $\s wz\tau$ as
$z\to 0$, that the apparent singularity of $R(\xi_{12})f$
at $z_1=z_2$ is removable.

Moreover it follows from \Ref{sig} that
$R(\xi_{12})f$ is 1-periodic in both variables and
that it has the required transformation properties under
translation by $\tau$ of both variables. From this we
deduce in particular that $R(\xi_{12})f$ is also
regular at $z_1=z_2+n+m\tau$, for all integers $n$ and
$m$ and is thus entire. \end{proof}

The translation operator $T_k(\xi)f(z)=f(z-\frac\xi k)$ maps
isomorphically $V_k=V_k(0)$ onto $V_k(\xi)$.
Let us define a modified $R$-operator as
\be
R_k(\xi_{12})
=T_k(\xi_1+\mu)^{-1}\otimes T_k(\xi_2)^{-1}
R(\xi_{12})
T_k(\xi_1)\otimes T_k(\xi_2+\mu)
\ee
It is defined for any complex $k$ and for positive
integer $k$ it preserves, by construction,
$V_k\otimes V_k$.
The notation is consistent, since  the right hand side is
indeed a function of the difference $\xi_{12}$,
 as a consequence of the elementary properties:

\begin{lemma}\label{ele}
{\rm (i)} $T_k(\xi+\eta)=T_k(\xi)T_k(\eta)$

{\rm (ii)} $R(\xi)$ commutes with $T_k(\eta)\otimes T_k(\eta)$
for any $\eta$.
\end{lemma}
More explicitly, $R_k$ is the operator
\bea
R_k(\xi)f(z_1,z_2)
&=&
\s \mu {z_{12}+\frac{\mu+\xi}{k}}\tau f(z_1+{\mu\over k},z_2-{\mu
\over k})\\
&-&
\s \xi {z_{12}+{\mu+\xi\over k}}\tau
f(z_2-{\xi\over k},z_1+{\xi\over k})
\eea
\begin{thm}
$R_k$ preserves $V_k\otimes V_k$ and
obeys the quantum Yang--Baxter equation.
\end{thm}

This  theorem is proved by reducing the QYBE for $R_k$
 to the QYBE for
$R$, using the rules of Lemma \ref{ele},

We thus get for any positive integer $k$ a quantum $R$-matrix
in End($V_k\otimes V_k$). We now identify this matrix, by computing
its matrix elements. The main technical tool here is the
action of the Heisenberg group $H_k$ on theta functions. The
group $H_k$ is generated by $A$, $B$ and a central element $\varepsilon$
subject to the relations
\be
A^k=B^k=1,\qquad AB=\varepsilon BA
\ee
It is well-known that theta functions of degree $k$
provide an irreducible representation of $H_k$ with
$\varepsilon=\exp(2\pi i/k)$. The action of generators
on $V_k$ is
\bea
Af(z)&=&f(z+{1\over k})\\
Bf(z)&=&e^{2\pi iz+\pi i\tau/k}f(z+{\tau\over k})
\eea
Diagonalizing the action of $B$ leads us to introduce
the functions
\bea
\theta_\alpha(z)=\sum_{n\in\Z}e^{\pi in^2\tau/k+2\pi in(z-\alpha/k)},
\qquad \alpha\in \Z_k.
\eea
These functions build a basis of $V_k$ and obey
\bea
A\theta_\alpha=\theta_{\alpha-1},\qquad B\theta_\alpha=
e^{2\pi i{\alpha/k}}\theta_\alpha
\eea
\begin{thm}
Define matrix elements of $R_k$ in the basis $\{\theta_\alpha\}$
of $V_k$ by
the formula
\be
R_k(\xi)\theta_\alpha\otimes\theta_\beta
=\sum_{\gamma,\delta\in\Z_k}
R_k(\xi)_{\alpha,\beta}^{\gamma,\delta}\theta_\gamma\otimes\theta_\delta
\ee
Then $R_k(\xi)_{\alpha,\beta}^{\gamma,\delta}$ vanishes unless
$\alpha+\beta=\gamma+\delta$, and if $\alpha+\beta=\gamma+\delta$,
\be
R_k(\xi)_{\alpha,\beta}^{\gamma,\delta}
=
{\vartheta_1({\mu-\xi-\alpha+\beta\over k},{\tau\over k})
\vartheta_1'(0,{\tau\over k})
\over
k\vartheta_1({\mu-\alpha+\gamma\over k},{\tau\over k})\vartheta_1
({\xi-\beta+\gamma\over k},{\tau\over k})}
\ee
\end{thm}
Thus the restriction of $R_k$ to $V_k\otimes V_k$ is
proportional to Belavin's solution \cite{B}, \cite{Ch},
\cite{Bo} (see \cite{RT} where the matrix elements are computed).
For $k=2$ it reduces
to Baxter's $R$-matrix of the eight-vertex model.
This results extends similar results obtained by Shibukawa
and Ueno \cite{SU} in the rational and trigonometric case.

To prove this theorem, notice first that $R_k(\xi)$ commutes
with $B\otimes B$ (and, in fact, also with $A\otimes A$). Thus
$R_k(\xi)\theta_\alpha\otimes\theta_\beta$ is a
linear combination of $\theta_\gamma\otimes\theta_\delta$
with $\gamma+\delta=\alpha+\beta$ (and its matrix elements
depend only on the differences of the indices).
Next, we need
to study the behaviour of $R_k(\xi)\theta_\alpha\otimes\theta_\beta$
under the action of $B\otimes{\rm Id}$. We use for this
the following decomposition of the function $\s wz\tau$
into eigenvectors for the translation by $\tau/k$.

\begin{lemma}
\be\sigma_w(z,\tau)={1\over k}\sum_{\gamma=0}^{k-1}
\s {(w+\gamma)/k}z{\tau/k}.\ee
\end{lemma}
\begin{proof}
Both sides of this equation have multipliers 1 and
$\exp2\pi iw$ as $z$ goes to $z+1$ and $z+\tau$.
Let us compare the poles. The right hand side has
possible poles on the lattice $\Z+k^{-1}\tau\Z$. The residue
at the pole $n+m\tau$ of the function (of $z$)
$\s {(w+\gamma)/k}z{\tau/k}$ is $\exp(2\pi i(w+\gamma)m/k)$
(see \Ref{sig}). By summing over $\gamma$, we see that the
residue vanishes at $n+m\tau/k$ unless $m$ is a multiple of $k$,
and is one at the origin. It follows that the difference
between the two sides of the equation is an entire function
of $z$ with multipliers 1 and  $\exp2\pi iw$  and must thus
vanish for generic $w$, and thus for
all $w$ by analyticity.
\end{proof}

We can now rewrite $R_k(\xi)\theta_\alpha\otimes\theta_\beta(z_1,z_2)$
as
\bea
\frac 1k&\sum_{\gamma\in\Z_k}
&
\{
\s {\mu+\gamma-\alpha\over k}{z_{12}+{\mu+\xi\over k}}{\tau\over k}
\theta_\alpha(z_1+{\mu\over k})
\theta_\beta(z_2-{\mu\over k})
\\
&&-
\s {\xi+\gamma-\beta\over k}{z_{12}+{\mu+\xi\over k}}{\tau\over k}
\theta_\alpha(z_2-{\xi\over k})
\theta_\beta(z_1+{\xi\over k})\}
\eea
Each summand $S_\gamma$ in this sum is an eigenvector of $B\otimes{\rm
Id}$, the eigenvalue being $\exp2\pi i\gamma/k$.  Thus this summand is
proportional to $\theta_\gamma\otimes\theta_\delta$ with
$\delta=\alpha+\beta-\gamma$.  To find the proportionality factor it
is sufficient to compute the summand $S_\gamma$ at any chosen point
(at which it does not vanish). If we chose $z_1$ and $z_2$ in such a
way that $z_1-z_2=(-\xi+\gamma-\alpha)/k$, then the
first term in $S_\gamma$
vanishes (since $\s ww\tau=0$) and we have
\be
\theta_\alpha(z_2-{\xi\over k})\theta_\beta(z_1+{\xi\over k})
=
\theta_\gamma(z_1)\theta_\delta(z_2)
\ee
It thus follows immediately that
\be
R_k(\xi)_{\alpha,\beta}^{\gamma,\delta}
=-{1\over k} \s {\xi+\gamma-\beta\over k}{\gamma-\alpha+\mu\over k}
{\tau\over k},
\ee
which, by \Ref{sigthe}, is what had to be shown.

\noindent{\it Acknowledgments.} Most of this work was done
as the first author was visiting IHES, which he thanks for
hospitality.

\end{document}